\documentclass[a4paper]{article}
\usepackage{hyperref}

\usepackage{INTERSPEECH2022}
\usepackage{url}
\usepackage{float}
\usepackage{comment}

\title{The Sounds of Home: A Speech-Removed Residential Audio Dataset for Sound Event Detection}
\name{Gabriel Bibbó, Thomas Deacon, Arshdeep Singh, Mark D. Plumbley}

\address{
  Centre for Vision, Speech and Signal Processing, University of Surrey, UK}
\email{\{g.bibbo,t.deacon,arshdeep.singh,m.plumbley\}@surrey.ac.uk}

\begin{document}

\maketitle

\begin{abstract}

This paper presents a residential audio dataset to support sound event detection research for smart home applications aimed at promoting wellbeing for older adults. The dataset is constructed by deploying audio recording systems in the homes of 8 participants aged 55-80 years for a 7-day period. Acoustic characteristics are documented through detailed floor plans and construction material information to enable replication of the recording environments for AI model deployment. A novel automated speech removal pipeline is developed, using pre-trained audio neural networks to detect and remove segments containing spoken voice, while preserving segments containing other sound events. The resulting dataset consists of privacy-compliant audio recordings that accurately capture the soundscapes and activities of daily living within residential spaces. The paper details the dataset creation methodology, the speech removal pipeline utilizing cascaded model architectures, and an analysis of the vocal label distribution to validate the speech removal process. This dataset enables the development and benchmarking of sound event detection models tailored specifically for in-home applications.

\end{abstract}
\noindent\textbf{Index Terms}: domestic soundscapes, wellbeing, older adults, residential audio dataset, sound event detection, speech removal.

\section{Introduction}






The increasing prevalence of smart home systems presents an opportunity to enhance the wellbeing of older adults through advanced audio machine learning (ML) technologies. 
Emerging AI technologies offer promising avenues for enhancing home wellbeing through sound monitoring, control, and generation \cite{ai4sEnvisioningSoundSensing2022}. However, a major obstacle for audio AI methods is the need for substantial labeled data tailored to specific tasks \cite{mesarosSoundEventDetection2021}.
A crucial component in this endeavor is the development of a comprehensive audio dataset that captures the complexities of Activities of Daily Living (ADL) within home environments. 
This paper addresses the need for such datasets, particularly focusing on older adults, to support the design and deployment of AI sound technologies aimed at improving quality of life.

The primary challenge in this domain is the collection of large volumes of data suitable for deep learning applications. The data must accurately reflect the context of home environments, necessitating long-duration recordings to capture a wide range of activities. However, this approach introduces significant privacy and data governance concerns, particularly in light of regulations such as General Data Protection Regulation (GDPR) \cite{gdpr2016}. Additionally, the data recording process must be unobtrusive to avoid disrupting daily life, requiring minimal intervention beyond initial installation and eventual removal.

To address these challenges, we developed a research study and data collection solution involving in-home audio capture with subsequent speech removal to mitigate privacy concerns. The audio dataset gathered focused on older adults in their homes to support people-centered design of AI sound technologies. 
The broader study ``Sound Wellbeing in Later Life'' employed a mixed methods approach \cite{deacon2023sound} to explore the domestic soundscapes of older adults (8 participants across 7 households). Key components included Co-Design Activities, Technology-assisted Deep-Listening Practice, and the deployment of an Audio Recording System. Our methodology showcases the mixed methods co-design of AI services for wellbeing within a Living Lab context. 

In this paper, we will focus on aspects related to the Audio Recording System (ARS). This system allowed us to capture indoor soundscapes in the homes of older people, creating a 1342 hours dataset of audio files for machine learning tasks and advancements in speech removal techniques.

The resulting dataset of audio files capturing indoor soundscapes can be utilized to train machine learning algorithms. These innovations can lead to more sophisticated sound sensing technologies that are sensitive to the nuances of indoor sound environments and their impact on wellbeing.
The key contributions of the paper are:
\begin{itemize}
\item We release a large multi-recording audio dataset recorded at multiple residential locations of older adults. This dataset includes 1342 hours of audio recordings and is intended for sound event detection research\footnote{The complete dataset is downloadable from the CVSSP server: \href{https://www.cvssp.org/data/ai4s/sounds_of_home}{\texttt{https://www.cvssp.org/data/ai4s/sounds\_of\_home}}. For segmented downloads, the dataset is also available in four parts on Zenodo. The first part is at \href{https://doi.org/10.5281/zenodo.12737915}{\texttt{https://doi.org/10.5281/zenodo.12737915}}, where links to the remaining parts are provided.}. 
\item To address privacy concerns within the recorded dataset, we implemented a data anonymization process that removes all speech and conversational content deemed to be personal information, designed to ensure compliance with data protection regulations.
\end{itemize}

\section{Dataset Collection Methodology}

An objective of this study was to collect a real-world audio dataset from domestic environments to support research on sound event detection in everyday conditions. To achieve this, we collaborated with the Living and Care Lab (LiCalab) at Thomas More University in Geel/Turnhout, Belgium, funded by the VITALISE Transnational Access programme \cite{licalab2024}.

\subsection{Ethics}
Ethical considerations were a priority throughout the study. 
The dataset collected was part of the ``Sound Wellbeing in Later Life'' study which received ethical approval via LiCalab (SMEC: G-2023 11 2174) and the University of Surrey (EGA ref: FEPS 23-24 001 EGA). 
As part of the ethical design of this study, we recognized the potential privacy concerns associated with recording audio data in domestic environments and implemented measures to protect participants' privacy and ensure responsible handling of the collected data. 
We obtained informed consent from all participants, ensuring they were fully aware of the study's purpose, the data collection process, and their right to withdraw at any time. 
As part of a risk assessment, we carefully considered the placement of the recording devices to minimize intrusion and prioritize participants' comfort and privacy within their homes.
To comply with GDPR, we implemented robust anonymization procedures to ensure no identifiable voice data would be included in the final dataset. This involved an automated speech removal process, followed by human verification, to eliminate any speech content present in the recordings. The technical details of our speech removal implementation will be discussed in the following section. The original audio data containing speech was then erased, and only the processed dataset without speech was retained for further analysis and archiving.

\begin{figure}[t]
  \centering
  \includegraphics[width=\linewidth]{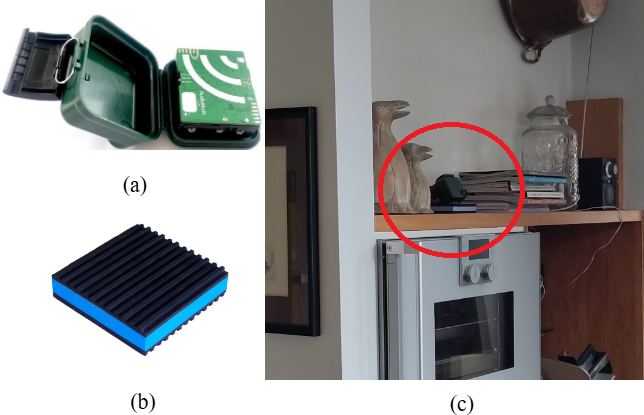}
  \caption{(a) AudioMoth, (b) Vibration pad, (c) AudioMoth in-situ on top vibration pad.}
  \label{fig:am-home}
\end{figure}

\subsection{Participants}

A key challenge was identifying an appropriate target population for the study. We collaborated closely with the Living and Care Lab (LiCalab) at Thomas More University in Geel/Turnhout, Belgium. LiCalab \cite{licalab2024}, a living lab specializing in working with older adults, has access to a panel of 800 older adults in the local area. This collaboration was crucial for recruiting suitable participants.

We focused on healthy individuals aged 55-80 years, including both retired and working adults nearing retirement. This decision was based on two factors: this group likely spends significant time at home engaging in various activities, and targeting healthy individuals minimized potential barriers during data collection.

We recruited 8 participants from urban and suburban areas around Geel/Turnhout. No specific criteria were set for dwelling types or technology affinity, aiming to capture diverse domestic environments and user profiles.

\subsection{Data Collection}

\begin{table*}[t]
    \caption{Data Collection Summary}
    \centering
    \small  
    \resizebox{1\textwidth}{!}{
        \begin{tabular}{p{0.25\textwidth}|p{0.65\textwidth}} \toprule
            \textbf{Aspect} & \textbf{Details} \\ \midrule
            Recording Device & AudioMoth with IPX7 Waterproof Case \\
            Number of Samples & 1344 audio files \\
            Length of Each File & 3595 seconds (approximately 1 hour) \\
            Number of Homes & 7 \\
            Recording Duration & 7 days per home \\
            Recording Window & Typically 8:00 AM to 9:00 PM \\
            Housing Types & Detached, semi-detached, mixed-use housing, high-rise flats \\
            Location Types & Rural and suburban areas \\
            Geography & Belgium (around Geel/Turnhout) \\
            Demographics & 8 participants, all aged 55-80 years \\
            Rooms Recorded & Primarily living rooms and kitchens \\
            Floor Types & Varied: wood, linoleum, concrete \\
            Wall Types & Varied: concrete, painted \\
            Home Features Diversity & Large windows, soft furnishings, open plan designs, pets (dog and cat in one home), robot vacuum cleaner (in one home) \\ \hline
            Special Notes & Two participants lived together in one house; Some homes had combined kitchen and living areas; Various noise issues noted (e.g., ventilation, ground-borne vibration) \\
            \bottomrule  
        \end{tabular}}
    \label{tab:data_collection_summary}
\end{table*}

The data collection process involved deploying two Audio Recording Systems (ARS) in each participant's home for a period of 7 days.  
The ARS consisted of two battery-powered acoustic sensors. We purchased and tested a total of four different audio recorders to find the most suitable device for our requirements. Those were: TASCAM DR-05X, TASCAM DR-40X, WILDLIFE ACOUSTICS Song Meter Mini, AudioMoth. Our selection criteria focused on user safety, audio quality, and device durability. We conducted experimental tests subjecting various devices to stress conditions to identify a configuration that preserved user safety and electrical integrity while minimizing transmission of low frequencies from furniture. After evaluating performance, waterproof capabilities, and ability to capture high-quality audio without significant deterioration from protective enclosures, we selected the AudioMoth devices \cite{AudioMoth2024} (see Figure \ref{fig:am-home} (a)). AudioMoths feature an analog MEMS microphone and were enclosed in an IPX7 Waterproof Case, resistant up to 1.5 meters under water. This setup met our criteria based on the experimental results. The optimal configuration is shown under the pad in Figure \ref{fig:am-home}(c) and Figure \ref{fig:am-home}(b).

The AudioMoths were configured to record audio in WAV (Waveform Audio Format) at a sample rate of 48 kHz for recordings without quality loss. The recordings duration was set to 3,595 seconds (approximately 1 hour) followed by a sleep period of 5 seconds, necessary for the device to save the file to the SD memory and start a new recording. This cycle was repeated continuously during the specified recording window, typically from 8:00 AM to 9:00 PM, to capture the soundscapes during the participants' most active periods.


To capture daily activities, one device was typically placed in the living room, and the other in the kitchen area, at a height between 1 and 2 meters, away from windows and obvious noise sources, such as motors or televisions. The specific placement of the devices was discussed with the participants to ensure their comfort and minimize interference with their daily routines. We also placed vibration pads under each AudioMoth to reduce impact of any ground-bourne vibration on audio recordings. A data collection summary can be seen in Table \ref{tab:data_collection_summary}.

To document the acoustic characteristics of the environments, we created detailed floor plans of each home, including precise measurements. Information about the construction materials used in the homes was also recorded, with the goal of reproducing the acoustic properties of the spaces in future research. After installation, photographs were taken of the recording devices in their respective placements for documentation purposes. The devices were installed by us, with assistance from LiCalab staff \cite{licalab2024}, to ensure proper setup and minimize potential disturbances to the participants' daily routines.

All recorded audio data was transferred securely from the SD memory cards on the AudioMoths to a backup hard drive by LiCalab staff immediately after collection. For further processing, the data was then transferred to the University of Surrey's SharePoint drive, that complies with the University's information security policies. Once the data transfer was confirmed, the SD card data was erased.


\section{Dataset Anonymisation Methodology}

To preserve privacy of participants, we perform post processing on recorded audios. The post-processing  involves using a Sound tagging model to identify speech information and then removing speech information from the recorded audios. Below, we explain the model selected for our study and a pipeline to remove speech.

\subsection{AI models for inference}
\label{aimodels}

To detect the presence of speech, we employed Pretrained Audio Neural Networks (PANNs) \cite{kong2020panns}. PANNs are neural networks pre-trained on the AudioSet dataset \cite{gemmeke2017audio}, which contains 2,084,320 YouTube videos spanning over 527 sound event classes, allowing the network to learn representations of diverse audio events. AudioSet contains 1,010,480 videos (48\%) labeled as containing speech. This proportion of speech data enables the pre-trained models to have a strong foundation for accurately detecting speech-related events \cite{kong2020panns}.
For this study, we employed two specific PANNs models: \textit{Cnn14\_DecisionLevelAtt} and \textit{Cnn14\_DecisionLevelMax}, chosen due to their proven performance in frame-wise analysis, which allows to identify the start time and duration of the sound events. On the AudioSet tagging task \cite{gemmeke2017audio}, the \textit{Cnn14\_DecisionLevelAtt} model achieved a mean Average Precision (mAP) of 0.425, while the \textit{Cnn14\_DecisionLevelMax} model obtained an mAP of 0.385 when evaluated on the whole dataset, and an Average Precision slightly above 0.8 for the Speech class \cite{kong2020panns}.

The \textit{Cnn14} architecture consists of 14 layers, including 6 convolutional layers inspired by VGG-like CNNs \cite{hershey2017cnn}. Each convolutional block contains two 3×3 convolutional layers, followed by batch normalization and ReLU activation to enhance training stability and speed. Average pooling with a 2×2 kernel is applied after each convolutional block for downsampling, as it has been shown to outperform max pooling. 

The architecture for both \textit{Cnn14\_DecisionLevelAtt} and \textit{Cnn14\_DecisionLevelMax} includes a global pooling layer after the last convolutional layer to summarize the feature maps into a fixed-length vector. The difference between the two models lies in the pooling mechanism: the \textit{Cnn14\_DecisionLevelAtt} model employs an attention mechanism to weigh the importance of different frame-level features, while the \textit{Cnn14\_DecisionLevelMax} model uses max pooling to select the most prominent features. 

After global pooling, an extra fully-connected layer is added to enhance the representation capability of the extracted features. Finally, a sigmoid activation function is applied to produce the presence probabilities for 527 sound classes. The training of these models utilizes binary cross-entropy loss to optimize the network parameters, ensuring accurate tagging and classification of audio events.

\subsection{Feature Extraction} 

The process of cleaning audio files in our study involves multiple steps to detect and remove speech segments while preserving other sound events. This workflow is conducted as follows. 

The audio file is processed by extracting log-mel spectrogram features, using a frame size (window size) of 1024 samples with a hop size of 512 samples, creating overlapping frames for analysis. Given the sampling rate of 48 kHz, each frame represents approximately 21.3 milliseconds of audio. A Hamming window is applied to reduce spectral leakage. The models use sequences of 32 frames at a time to perform inference, that was explained in more detail in Section \ref{aimodels}. 

The detection results, including frame indices, probabilities, and metadata, are saved in JavaScript Object Notation (JSON) format, that is a standard text-based format for representing structured data based on JavaScript object syntax. The JSON files detail the seven most significant sound events detected in each frame, providing a confidence score for each event label ranging from 0 to 1 probability. 

\subsection{Threshold Analysis} 

\begin{figure}[H]
  \centering
  \includegraphics[scale=0.6]{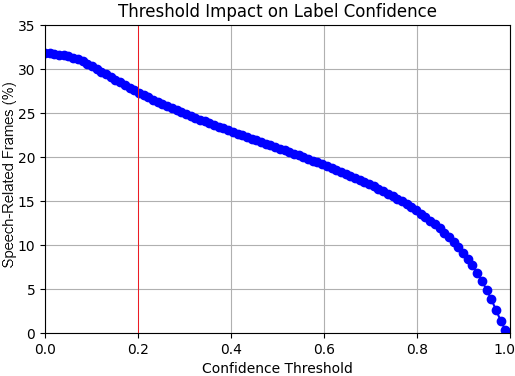}
  \caption{
Plot of the proportion of the entire audio dataset where any of the predefined vocal-related labels exceed given confidence thresholds.}
  \label{fig:threshold}
\end{figure}

The audio dataset analysis focused on the distribution of predefined vocal-related labels to determine an optimal confidence threshold that could effectively differentiate spoken voice from ambient sounds while preserving as much of the dataset as possible. The goal was to remove all audio segments containing speech-related labels without significantly reducing the total hours of audio data. To cover all situations in which the audio could contain spoken voice potentially identifiable as personal information, all segments above the threshold were removed that included the labels: ``Speech'', ``Singing'', ``Male singing'', ``Female singing'', ``Child singing'', ``Male speech, man speaking'', ``Female speech, woman speaking'', ``Conversation'', ``Narration, monologue'', ``Music''. Figure \ref{fig:threshold} shows the 

\noindent relationship between various confidence thresholds and the proportion of the dataset containing the predefined vocal-related labels with confidence values exceeding these thresholds.  

The graph \ref{fig:threshold} starts at 32\% at a zero threshold instead of 100\% due to data compression in our annotation process. PANNs produces a vector of 527 classes; however, to reduce the size of the JSON annotation files, we retained only the top 7 most relevant classes. If all 527 classes had been stored, any frame would likely exhibit a non-zero confidence for vocal-related labels at a zero threshold, representing 100\% of the dataset. The omission of residual classes that do not provide significant information results in 32\% of frames displaying vocal-related labels at the zero threshold. 

A threshold was set at 0.2 based on both the aforementioned statistical analysis and manual verification, involving listening sessions to ensure the absence of spoken voice below the threshold and its presence above it. This process confirmed the threshold's adequacy in excluding spoken content while retaining relevant non-vocal audio data, essential for maintaining dataset integrity in sound event detection studies. 

\subsection{Speech Removal Process} 

A subsequent cleaning phase involves reading the JSON files and processing the audio files to remove the detected speech segments. A 0.2 probability threshold is applied to confidence values to determine speech presence. The script calculates the start and end times for each detected speech segment, extending these intervals by one second before and after the detected speech for thorough removal.

\begin{figure}[H]
  \centering
  \includegraphics[scale=0.37]{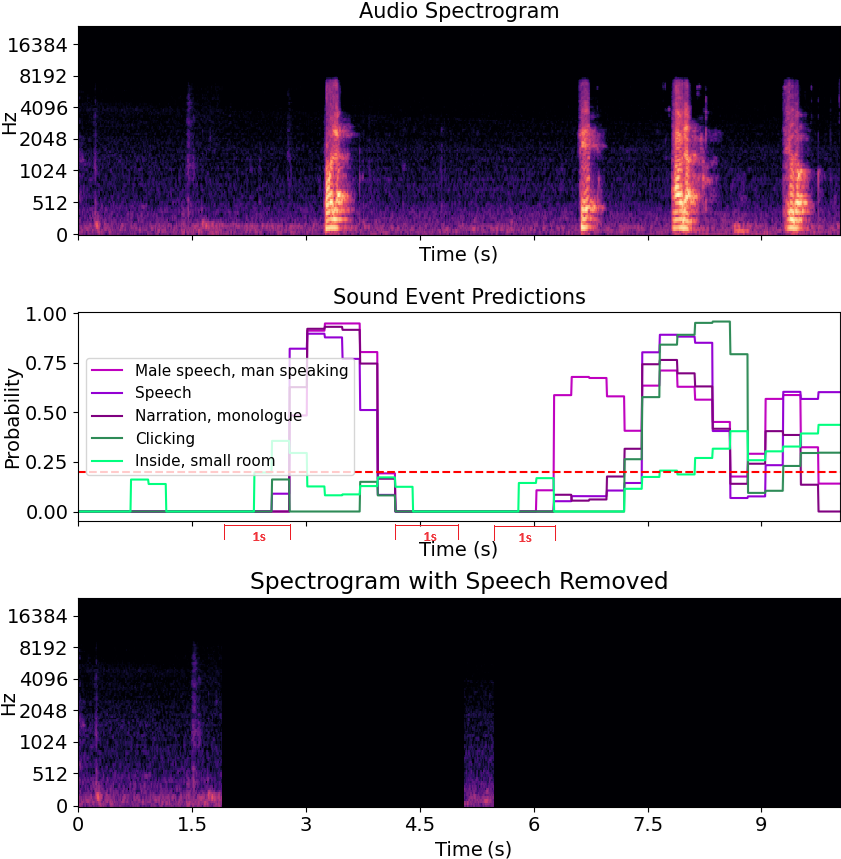}
  \caption{
Top plot shows the audio spectrogram of male speech. The middle plot displays probability predictions of the top 5 sound events using PANNs. The bottom plot presents the spectrogram after removing voice segments exceeding a 0.2 probability threshold, extended by one second before and after for complete removal.}
  \label{fig:combined_plot}
\end{figure}

The removal of speech modifies sections of the audio file containing speech by replacing these with low-amplitude noise, generated at $10^{-10}$ relative amplitude. This step prevents division-by-zero errors during subsequent processing and maintains the audio file's structural integrity. The modified file, now free of speech, is saved to a new location in the temporary storage cluster server. Figure \ref{fig:combined_plot} illustrates this process on a 10-second audio fragment with spoken voice.

To ensure thorough removal of speech, a two-step process is used with two models in cascade. Initially, the audio file is

\noindent processed, creating an intermediate file on the cluster server. This file undergoes a second processing with a different model, then is deleted to maintain a clean workspace.


%

\section{Conclusions}

We have introduced \textit{The Sounds of Home} dataset of domestic speech-removed audio events.
The objective of the dataset is to produce a privacy-compliant dataset of sounds captured by an edge device deployed in the field.
In this paper we have reported the steps taken to achieving privacy compliance, demonstrating a need for future research in this area. As technology advances into sensitive domains like elder care, balancing innovation with privacy protection becomes increasingly complex, necessitating ongoing research to develop effective, ethical solutions.
We are releasing this data to accelerate research in the area of acoustic event detection for research into domestic settings and assistive technologies for older adults at home. Our GitHub repository \cite{bibbo2022voice} provides the link to the dataset hosted at Zenodo.
In the future, we plan to create a ground truth of manual annotations by using our AI-generated annotations to assist multiple human annotators, each working on dataset subsections. Additionally, we plan to utilize the redundant information from the two recorders placed simultaneously in the homes, along with the house floor plans, to recreate the acoustic properties of the households.

\section{Acknowledgements}

This work was supported by Engineering and Physical Sciences Research Council (EPSRC) Grant EP/T019751/1 ``AI for Sound (AI4S)''. For the purpose of open access, the authors have applied a Creative Commons Attribution (CC BY) licence to any Author Accepted Manuscript version arising.

\bibliographystyle{IEEEtran}

\bibliography{mybib}

\end{document}